\documentclass[preprint,12pt,onecolumn]{elsarticle}
\usepackage[
left=3cm,
right=3cm,
top=2.5cm,
bottom=2.5cm
]{geometry}
\usepackage{lineno,siunitx,verbatim,xcolor}
\usepackage{pstricks,graphicx,amsmath,bbm,mathrsfs,amssymb,times,siunitx}
\journal{arXiv.org}
\begin{document}
\begin{frontmatter}
\nolinenumbers

\title{Low frequency-to-intensity noise conversion in a pulsed laser cavity locking by exploiting
Carrier-Envelope Offset manipulation}

\author[aff1,aff2]{F. Canella}
\author[aff3,aff2]{E. Suerra}
\author[aff2]{D. Giannotti}
\author[aff4,aff2]{G. Galzerano}
\author[aff3,aff2]{S. Cialdi}
            
\affiliation[aff1]{organization={Dipartimento di Fisica, Politecnico di Milano},
            addressline={Piazza Leonardo da Vinci 32}, 
            city={Milano},
            postcode={20133}, 
            state={Italy},
            country={}}
            
\affiliation[aff2]{organization={Istituto Nazionale di Fisica Nucleare, Sezione di Milano},
            addressline={via Celoria 16}, 
            city={Milano},
            postcode={20133}, 
            state={Italy},
            country={ }}
  
 \affiliation[aff3]{organization={Dipartimento di Fisica, Universita degli Studi di Milano},
            addressline={via Celoria 16}, 
            city={Milano},
            postcode={20133}, 
            state={Italy}}     
            
\affiliation[aff4]{organization={Istituto di Fotonica e Nanotecnologie - CNR},
            addressline={Piazza Leonardo da Vinci 32}, 
            city={Milano},
            postcode={20133}, 
            state={Italy},
            country={}}


\begin{abstract}
We report on the dependence of the frequency-to-intensity noise conversion in the locking of an ultrafast laser against a high-finesse optical resonator from the Carrier Envelope Offset (CEO) frequency.
By a proper combination of the cavity finesse and laser CEO frequency, it is possible to optimize the signal-to-noise ratio of the laser intensity trapped into the optical resonator.
The theoretical description of the problem together with the numerical simulations and experimental results are presented with the aim of a strong suppression of the intensity fluctuations of the trapped laser field.
\end{abstract}
\end{frontmatter}

\section{Introduction}
Frequency-to-intensity noise conversion in the laser stabilization with respect to high-finesse optical resonator is a well known problem \cite{Villar2008,Morville02,Ma99} and over the years several experimental solutions have been implemented to reduce the intensity noise of the cavity trapped radiation: the use of noise-immune high-frequency modulation detection schemes \cite{Ma99}, by means of wide control loop bandwidths in the frequency locking schemes \cite{gatti},  and using cavity ring-down methods \cite{berden10,Long2014}. In the majority of the investigated cases, the laser sources operated in a continuous wave regime. However, in the last twenty years thanks to the introduction of the optical frequency comb sources, the locking of ultrafast broadband laser sources to high-finesse resonators is of extreme interest in a wide variety of sensing/spectroscopic applications \cite{Adler,Bernhardt2010,Hog} as well as in extreme non-linear optics for the generation of coherent radiation in the UV and XUV spectral regions \cite{JonesPRL,Gohle2005,AllisonPRL}. In these applications, cavity locking is exploited to increase by several orders of magnitude either the interaction path with the spectroscopic samples or the circulating laser intensity. In any case, low intensity noise of the trapped laser field is a mandatory requirement to obtain the highest signal-to-noise ratio (SNR) and temporal coherence.

In this paper, we demonstrate a novel solution to reduce the residual frequency-to-intensity noise conversion when a pulsed laser is coupled to a high-finesse resonator.
This is related to the use of slightly detuned resonance conditions between the cavity resonances and laser modes by acting on the carrier envelope offset frequency of the pulsed laser source.
In this way, laser modes distant from the center of its spectrum contribute to the second order derivative frequency to intensity noise conversion with opposite sign with respect to the central modes and therefore their quadratic contributions cancel out.
By a proper combination of the CEO frequency detuning and cavity finesse values, large cavity gain factor can be obtained together with a strong reduction of the frequency-to-intensity noise conversion, an essential requirement for the generation of X-ray radiation by inverse Compton scattering (ICS) \cite{brixs} with low-intensity noise.
After the theoretical description of the method, a detailed experimental characterization of the technique is performed using a mode-locked Yb:fiber laser at \SI{1030}{\nano\meter}, and a ring cavity optical resonator.
With a gain factor as large as $1130$, we demonstrated an integrated frequency to intensity noise conversion reduced from \SI{8}{\percent} to less than \SI{0.5}{\percent}.

The paper is structured in four sections.
The principles and the theoretical treatment of the method is presented in Sec. \ref{sec:theory}, whereas the numerical simulation are reported in Sec. \ref{sec:simul}.
Sec. \ref{sec:experiment} shows the experimental characterization of the method.
Finally, Sec. \ref{sec:concl} closes the paper with some concluding remarks.

\section{Theory}\label{sec:theory}
The link between the CEO frequency $f_{\rm ceo}$ and the frequency-to-noise conversion in an optical cavity can be shown starting from the time-domain electric field of a mode-locked laser.
The train of pulses in time domain is given by the superposition of the field of several modes oscillating in phase.
If every mode has a frequency $\nu_{m}$, being $m \!\in\! \mathbb{N}$, the electric field $\mathscr{E}\!\left(t\right)$ can be written as:
\begin{equation}\label{eq:pulsedlaser}
 \mathscr{E}\!\left(t\right) = 
 \sum_{m=0}^{\infty}
\sqrt{S_m} \,
e^{ -i 2\pi\nu_m t + i \varphi_m\!\left(t\right) }
\end{equation}
where $S_m$ is the laser power spectrum $S\left(\nu\right)$ evaluated at the frequency $\nu_m$, while $\varphi_m$ is a generic phase noise.
Typically, $\varphi_m$ contains both slow components (essentially related to mechanical vibrations), and fast components.
In the spectral domain, Eq. \ref{eq:pulsedlaser} becomes:
\begin{align}\label{eq:lasercombspectral}
E\!\left(\nu\right) = 
\mathfrak{F}\left[\mathscr{E}\!\left(t\right)\right] = 
\sum_{m=0}^{\infty}
\sqrt{S_m}
\int e^{i 2\pi\left(\nu - \nu_m\right)t}
e^{i \varphi_m\left(t\right)} \, {\rm d}t
\end{align}
where $\mathfrak{F}\left[\cdot\right]$ denotes the Fourier transform.
When $\Delta\nu_m \ll \nu_0$, we assume that the noise $\varphi_m\!\left(t\right) = \varphi\!\left(t\right)$ is equal for all the laser teeth.
%
%
Furthermore, we assume that $\varphi\!\left(t\right) \ll 2\pi$, so that we can expand the exponential to first order, leading to
\begin{align}\label{eq:lasercombspectral2}
E\!\left(\nu\right) &= 
 \sum_{m=0}^{\infty}
\sqrt{S_m}
\int e^{i 2\pi\left(\nu - \nu_m\right)t}
\left(1 + i \,\varphi\left(t\right)\right) \, {\rm d}t = \nonumber \\
&= \sum_{m=0}^{\infty}
\sqrt{S_m}
\left( \delta\left(\nu-\nu_m\right) + 
i \, \phi\!\left(\nu-\nu_m\right)
\right) 
\equiv
\sum_{m=0}^{\infty} E_m\!\left(\nu\right)
\end{align}
Equation \ref{eq:lasercombspectral2} tells us that the laser spectrum has a comb-like structure of Dirac deltas, broadened by the noise $\phi\!\left(\nu\right) = \mathfrak{F}\left[\varphi\!\left(t\right)\right]$.
Thus, the laser power is given by:
\begin{align}
\label{eq:Plaser}
{\rm P}^{\rm {(laser)}}
=
\sum_{m=0}^{\infty}
S_{m}
\left| \delta\left(\nu-\nu_m\right) + 
i \, \phi\!\left(\nu-\nu_m \right) \right|^2
=
\sum_{m=0}^{\infty}
S_{m}
\, \Phi_m
\equiv
\sum_{m=0}^{\infty}
P^{\rm {(laser)}}_{m}
\end{align}
where $\Phi_m = \left| \delta\left(\nu-\nu_m \right) + i\,\phi\left(\nu-\nu_m \right) \right|^2$.\\
Note that in Eq. \ref{eq:Plaser} the crossed products cancel since different modes do not overlap.

To complete the discussion about the laser, we remind that the frequencies of its modes are  \cite{picque2019}
\begin{equation}\label{eq:lasercombfreq}
\nu^{\rm (laser)}_m = 
m\, f_{\rm rep} + \frac{\Delta\phi_{\rm cep}}{2\pi}f_{\rm rep} = 
m\, f_{\rm rep} + f_{\rm ceo}
\end{equation}
where $f_{\rm rep}$ is the separation of the teeth corresponding to the repetition rate of the laser, while $\Delta\phi_{\rm cep}$ is its Carrier-Envelope Phase shift, generated by the different phase and group velocities in the laser cavity.
We also defined the Carrier-Envelope Offset as $f_{\rm ceo} = \frac{\Delta\phi_{\rm cep}}{2\pi}f_{\rm rep}$.

The laser pulses can be coupled to an optical cavity, in order to stack them and increase their power with a passive gain proportional to the resonator finesse.
In particular, for an overcoupled cavity of finesse $F$, the maximum achievable gain is $\frac{2}{\pi}F$.

The spatial modal structure of the cavity is given by the well known Hermite-Gaussian polynomial, with optical frequencies corresponding to
\begin{equation}\label{cavityfreq}
\nu_{n,x,y}^{\rm (cav)} = 
{\rm FSR} \left( n + 
\frac{x+1/2}{2\pi}\arccos{\left(M_{\rm H}\right)} + 
\frac{y+1/2}{2\pi}\arccos{\left(M_{\rm V}\right)}
\right)    
\end{equation}
where $n$, $x$ and $y$ are positive integers that indicate the longitudinal, horizontal and vertical order of the mode, respectively, while $\rm FSR$ is the Free Spectral Range of the cavity.
$M_{\rm H}$ and $M_{\rm V}$ are the horizontal and vertical stability parameters, respectively, found by the Round-Trip Matrix of the cavity in the ABCD-matrix formalism.
In general, $M_{\rm H}$ can be different from $M_{\rm V}$, and they must stay in the range $\pm1$ in a stable resonator \cite{Svelto2010}.
In general, $\rm FSR$ is inversely proportional to the cavity length and depends on the wavelength in case of significant intracavity and/or mirrors dispersion.
At this level, we neglect these dispersion effects.

\begin{figure}
\label{fig:combscoupling}
\centering
\includegraphics[width=0.8\textwidth]{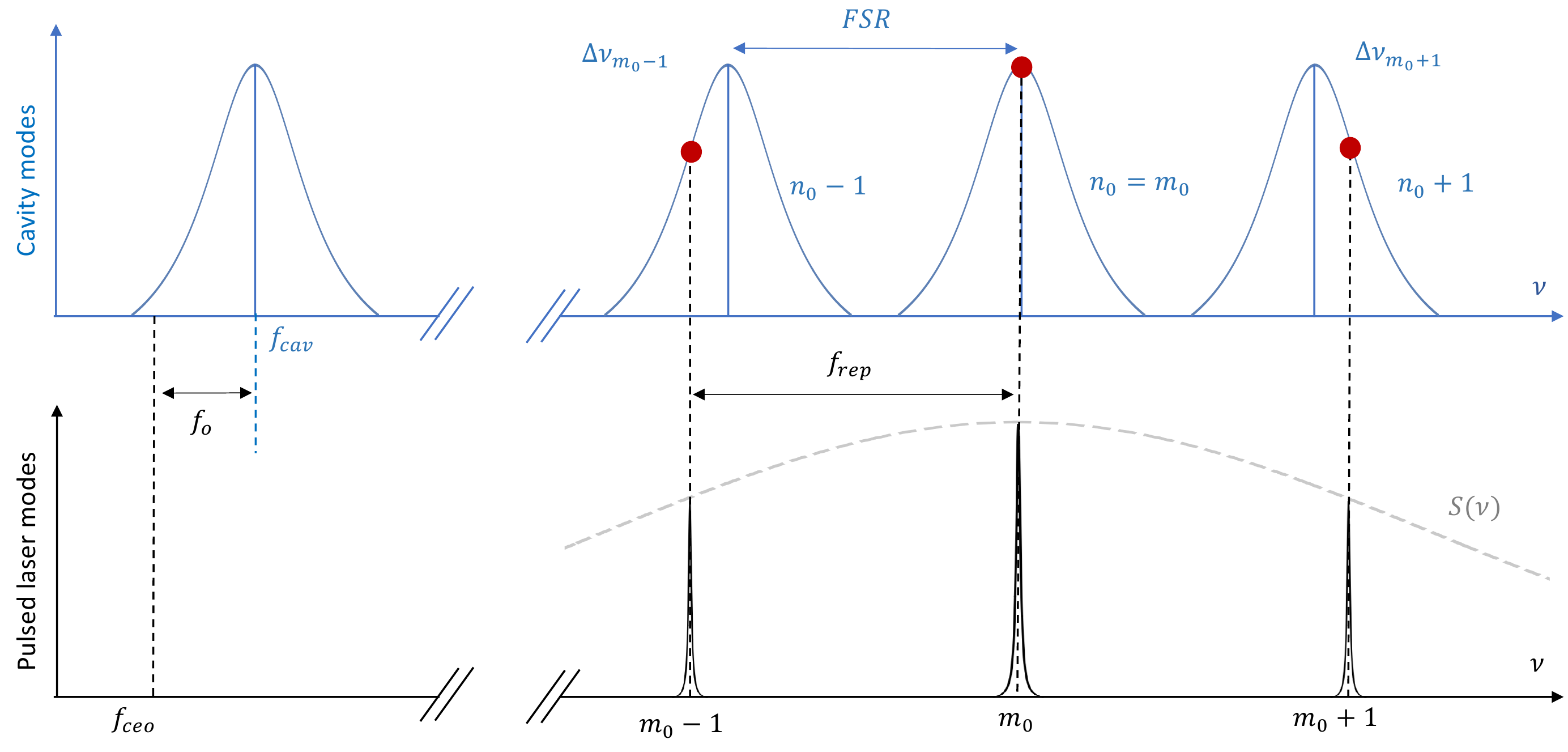}
\caption{Laser and cavity modes in frequency domain.
In this representation, the cavity is stabilized to the laser on the tooth $n_0 = m_0$ with the PDH technique. 
The teeth with indexes $n \neq m_0$ have a detuning $\Delta\nu_{m}$, which is positive for $m>m_0$ or negative for $m>m_0$.
In this scheme $f_{\rm cav} > f_{\rm ceo}$, so $f_{0} < 0$.}
\end{figure}

If we consider the fundamental mode only ($x=y=0$), Eq. \ref{cavityfreq} becomes
\begin{equation}\label{eq:cavitymodefreq}
\nu^{\rm (cav)}_n = 
n\, {\rm FSR} + f_{\rm cav}.    
\end{equation}
where we defined $f_{\rm cav} = {\rm FSR}\left[\arccos{\left(M_{\rm H}\right)}/4\pi + \arccos{\left(M_{\rm V}\right)}/4\pi\right]$.
The highest passive gain is achieved when the laser and the cavity teeth are perfectly coupled.
This condition can be obtained and maintained during time exploiting the well known Pound-Drever-Hall (PDH) technique \cite{blackpdh}, which, in our case, stabilizes the $\rm FSR$ of the cavity to match the laser teeth.
Starting from Eq. \ref{eq:lasercombfreq} and Eq. \ref{eq:cavitymodefreq}, if we stabilize the $n_0^{\rm th}$ tooth of the cavity with the $m_0^{\rm th}$ tooth of the laser, with $n_0 = m_0$, we can write
\begin{align*}
\nu^{\rm (laser)}_{m_0} = \nu^{\rm (cav)}_{m_0} \quad
\Rightarrow \quad
m_0 \, f_{\rm rep} + f_0 = m_0 \, {\rm FSR}
\end{align*}
where $f_0 = f_{\rm ceo} - f_{\rm cav}$ is the relative frequency offset between laser and cavity.
The PDH technique fixes the FSR of the cavity to ${\rm FSR} = f_{\rm rep} + \frac{f_0}{m_0}$.
A clear representation of the two combs coupling is reported in Fig. \ref{fig:combscoupling}.
A perfect overlap of the cavity and the laser modes is possible only if $f_0=0$, otherwise, each tooth has a detuning given by $\Delta\nu_{m} = \frac{f_0}{m_0} \Delta m$, where $\Delta m = m - m_0$.
When the laser-cavity locking condition is achieved, only relative detunings are relevant.
Thus, it is equivalent to introduce noise from the laser or the cavity, and from now on we attribute all the noise to the laser without loss of generality.
On the other hand, the same results can be achieved by assuming $\varphi\!\left(t\right)=0$, while introducing a noise $\delta{\rm FSR}\!\left(t\right)$ on the cavity frequencies.

Now, we study how $f_0$ affects the laser-cavity coupling, hence the intracavity power and its noise.
We start from considering the intracavity power in relation with the incoming laser power and the cavity response:
\begin{align}
\label{eq:Pm}
{\rm P}^{\rm (cav)} &=
\sum_{m=0}^{\infty}
P^{\rm (laser)}_m
\,\frac{1-R_1}{1 + R -2\sqrt{R} \cos{\frac{2\pi\Delta\nu_m}{\rm FSR}}} = \nonumber \\
&=
\sum_{m=0}^{\infty}
P^{\rm (laser)}_m
\,\Gamma\!\left(\Delta\nu_m\right)
\equiv
\sum_{m=0}^{\infty}
P^{\rm (cav)}_m\!\left(\Delta\nu_m\right)
\end{align}
where $\Gamma\!\left(\Delta\nu_m\right)$ is the gain of the cavity for the mode $m$ detuned by $\Delta\nu_m$, $R_1$ is the input cavity mirror power reflectivity, and $R$ the product of all the cavity mirrors' reflectivity.
The total cavity gain $\Gamma_{\rm tot}$ is then given by the weighted average of the single modes gains over all the coupled modes.
In terms of power, assuming an incoming radiation power $\rm P^{\rm (laser)}$, the stored power is simply given by $\rm P^{\rm (cav)} = P^{\rm (laser)} \cdot \Gamma_{\rm tot}$. 
\begin{align}
\Gamma_{\rm tot} = \frac{
\sum_{m} P^{\rm (cav)}_{m} \! \left(\Delta\nu_m\right) 
}{
\sum_{m} P^{\rm (laser)}_m 
}
\end{align}
While $\Gamma_{\rm tot}$ is not directly accessible, the transmitted power does, and it is directly proportional to the gain as $\rm P^{\rm (trans)} = \left(1 - R_2\right) P^{\rm (laser)}\Gamma_{\rm tot}$.

As far as the peak power of the intracavity pulses ${\rm P}_{\rm peak}$ is concerned, we find an implicit dependence on $f_0$ hidden in the coupling with the cavity.
Indeed, the cavity acts as a filter for the laser, both in amplitude and in phase.
The additional phase experienced by each mode, which impacts on the temporal structure of the pulses, must be taken into account to properly estimate ${\rm P_{\rm peak}}$.
As $f_0$ increases, both the cavity spectral filtering effect and the phase become more and more important, broadening in time the stored pulses and lowering ${\rm P}_{\rm peak}$.
We can write:
\begin{align}
\label{eq:Ppeak}
{\rm P_{\rm peak}} = \max{
\left\{ \left|\mathfrak{F}^{-1}\left[
\sqrt{S\left(\nu\right)}\,F_{\rm cav}\left(\nu\right)\, e^{i\phi_{\rm cav}\left(\nu\right)}
\right]\right|^2
\right\}
}
\end{align}
where $F_{\rm cav}$ and $\phi_{\rm cav}$ are the cavity filter function in amplitude and phase.
%
%
%
In experimental setups there is often an additional detuning $f_{\rm pdh}$, given by an electronic offset in the PDH stabilization, which locks the mode $m_0$ not exactly on the line maximum.
The PDH stabilization offset can be included in the theory by simply considering $\Delta\nu_m = \left( \frac{f_0}{m_0} \right)\Delta m + f_{\rm pdh}$.

\begin{figure}
\centering\includegraphics[width=0.9\textwidth]{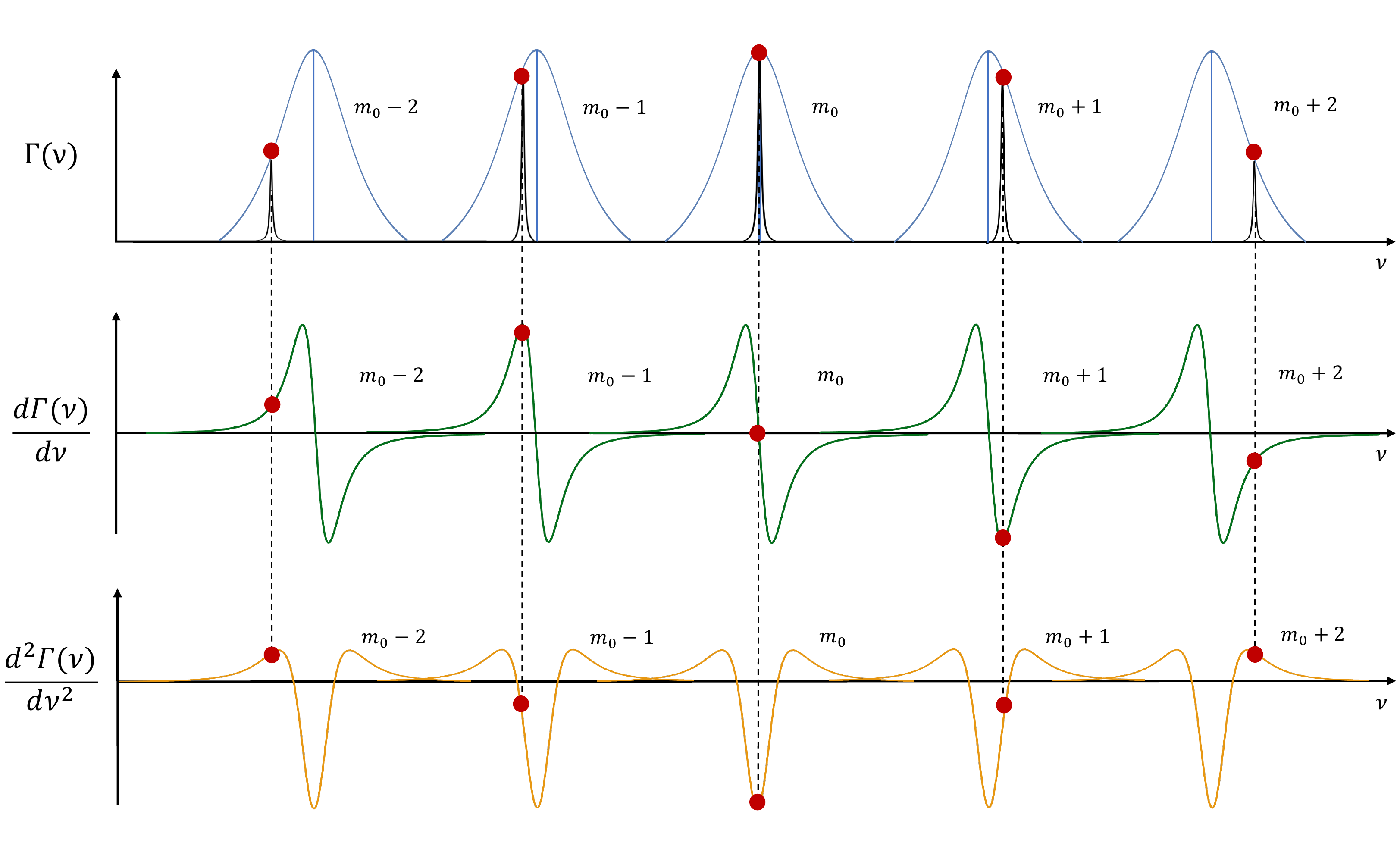}
\caption{Representation of the effect of $f_0$ on the evaluation point of the cavity gain and its derivatives (from above to below: $\Gamma$ with the laser teeth, $d\Gamma/d\nu$, and $d^2\Gamma/d\nu^2$ as functions of $\nu$).
At increasing offset, the gain progressively decreases with the distance from the central mode $m_0$.
The first derivative terms on the right balance the ones on the left, since $d\Gamma/d\nu$ is an even function (when $f_{\rm pdh}=0$).
On the other hand, $d^2\Gamma/d\nu^2$ is odd.
Thus, the central modes contributions are compensated by the external modes, where the second order derivative becomes positive.}
\label{T}
\end{figure}

At this point, we are able to debate on a particularly interesting and important issue: the noise transfer from the laser and the cavity to the stored power in the coupled system.
There are essentially two noise sources for the stored power: the laser intensity noise and the frequency noise.
Intensity noise is substantially due to power fluctuations of the laser source, and we will not cover it in this work.
On the contrary, frequency noise induces fluctuations in the stored power because it causes time-dependent additional detuning $\delta\rm{f}\!\left(t\right)$.
This frequency noise is strictly dependent on the phase noise, indeed $\delta\rm{f}(t)=(2\pi)^{-1} d\varphi(t)/dt$.
Furthermore, it can be associated to a $\delta f \!\left(\nu\right)$ and a spectral distribution with a standard deviation $\sigma_{\delta f}$.
The introduction of $\sigma_{\delta f}$ is quite important, because this is a parameter directly accessible from the experimental setup, by measuring the integrated frequency noise discriminated by the cavity while the locking with the laser is maintained \cite{coluccelli2015}.
To study the influence of $\delta f$ on the power fluctuations, the function $\Gamma\!\left(\Delta\nu_m+\delta f\right)$ can be expanded around the offset detuning frequency $\Delta\nu_m$ as
\begin{equation}
\label{sviluppoT}
\Gamma\!\left(\Delta\nu_m+\delta f\right) \approx \Gamma\!\left(\Delta\nu_m\right)+\frac{d\Gamma}{d\nu}\bigg|_{\Delta\nu_m}\delta f+\frac{1}{2}\frac{d^2\Gamma}{d\nu^2}\bigg|_{\Delta\nu_m}\delta f^2.
\end{equation}
Therefore, recalling that the total cavity gain is the average gain of the teeth, weighed on the coupled laser spectrum $S\left(\nu\right)$, and the same holds for the fluctuations, we separate the gain in two components $\Gamma_{\rm tot}$ and $\delta \Gamma_{\rm tot}$, respectively writable as:
\begin{equation}
\label{eq:sviluppoTtot}
\Gamma_{\rm tot} = \frac{ \sum_{m} S_m \Gamma\left(\Delta\nu_m+\delta f\right)}{\sum_{m} S_m}
\approx  \frac{ \sum_{m} S_m \Gamma\left(\Delta\nu_m\right)}{\sum_m S_m}
=  \frac1N \sum_{m} S_m \Gamma\left(\Delta\nu_m\right)
\end{equation}
\begin{align}
\label{svildeltaTtot}
\delta \Gamma_{\rm tot}\left(\delta f\right) =& \frac1N \sum_{m} S_m 
\left[ \Gamma\left(\Delta\nu_m+\delta f\right) - \Gamma\left(\Delta\nu_m\right)
\right] = \nonumber\\
=& \frac1N \left( \sum_{m} S_m 
\frac{d\Gamma}{d\nu}\bigg|_{\Delta\nu_m}\delta f
+\frac{1}{2}\sum_{m} S_m \frac{d^2\Gamma}{d\nu^2}\bigg|_{\Delta\nu_m} \delta f^2 \right) 
\end{align}
$\Gamma_{\rm tot}$ in Eq. \ref{eq:sviluppoTtot} is noise-immune, and it is only an implicit function of the system offset $f_0$ and of the eventual $f_{\rm pdh}$.
We also defined $N\equiv\sum_{m} S_m$ to simplify the notation on the latter equations.
The first term of Eq. \ref{svildeltaTtot} is an odd function when $f_{\rm pdh}=0$, thus the sum over all the modes around $m_0$ vanishes, because $\Delta\nu_m<0$ or $\Delta\nu_m>0$ for $m<m_0$ or $m>m_0$, respectively.
Hence $\frac{d\Gamma}{d\nu}\big|_{\Delta\nu_m}<0$ or $\frac{d\Gamma}{d\nu}\big|_{\Delta\nu_m}>0$.
In the case of $f_{\rm pdh}$ differs from zero, the symmetry is broken and a non-negligible noise from this term arises, although it that can be reduced by improving the stabilization.
The second term is the most important for us, because it can be adjusted by manipulating the offset $f_0$ (hence $f_{\rm ceo}$) to reduce the noise transfer.
In Fig. \ref{T} a representation of the noise transfer reduction via $f_0$ principle is given.
The second-order derivative of $\Gamma$ is an even function, with both positive and negative values encountered at increasing detuning.
When $f_0=0$, all the teeth contributes with a negative $\frac{d^2\Gamma}{d\nu^2}\big|_{\Delta\nu_m}$, so that the noise sums constructively.
On the other hand, when $f_0$ increases, the evaluation point of the derivative changes proportionally with $\Delta m$.
The most external modes in the spectrum can have $\frac{d^2\Gamma}{d\nu^2}\big|_{\Delta\nu_m}>0$, thus they compensate the parabolic noise contributions of the internal teeth close to $m_0$.
The more $f_0$ grows, the more the noise suppression is efficient, but the drawback is that an increasing of $f_0$ lowers $\Gamma_{\rm tot}$.
For this reason, the key-point is to compare the gain loss with the noise suppression effect.
As we will show in the next sections, the noise suppression is more effective than the gain drop, and, consequently, it is possible to dramatically reduce the intensity noise without a large decrease in the stored power.

A last step useful to understand the impact of noise suppression is to evaluate how a frequency noise $\sigma_{\delta f}$ affects the integrated noise $\sigma_{P}$ of the intracavity power.
This is directly given by the fluctuation of the gain $\Gamma_{\rm tot}$, namely $\sigma_{\Gamma}$.
To simplify the notation, we define $\alpha \equiv \frac1N \sum_{m} S_m \frac{d\Gamma}{d\nu}\big|_{\Delta\nu_m}$ and $\beta \equiv \frac{1}{2 N}\sum_{m} S_m \frac{d^2\Gamma}{d\nu^2}\big|_{\Delta\nu_m}$, then:
\begin{align}
\sigma_{\Gamma}^2  = &
\left<\delta \Gamma_{\rm tot}^2\right>-\left<\delta \Gamma_{\rm tot}\right>^2 =
\left<\left(\alpha\,\delta f + \beta\, \delta f^2\right)^2\right>
-\left<\alpha\,\delta f + \beta\, \delta f^2\right>^2 = \nonumber \\
= & \alpha^2\left<\delta f^2\right>
+\beta^2\left<\delta f^4\right>
+2\alpha\beta\left<\delta f^3\right>
-\alpha^2\left<\delta f\right>^2
-\beta^2\left<\delta f^2\right>^2
-2\alpha\beta\left<\delta f\right>\left<\delta f^2\right>= \nonumber \\
= & \alpha^2 \sigma_{\delta f}^2
+ \beta^2 \sigma_{{\delta f}^2}^2
\label{variance}
\end{align}
In case of no PDH offset, $\alpha=0$ and Eq. \ref{variance} becomes $\sigma_{\Gamma} = \left|\beta\right| \, \sigma_{{\delta f}^2}$.
Notice that in the calculations, we exploited the fact that $\left<\delta f\right> = 0$ and $\left<\delta f^3\right> = 0$ for symmetry.
The relative noise of the cavity gain is simply given by
\begin{align}
\sigma_{\Gamma,{\rm rel}} = \frac{\sigma_{\Gamma}}{\Gamma_{\rm tot}}
\end{align}
Equivalently, the relative intracavity power noise is $\sigma_{P,{\rm rel}} = \sigma_{P} / P_{\rm tot}^{\rm (cav)} =\sigma_{\Gamma,{\rm rel}}$.

Finding a simple relation between $\sigma_{\delta f}$ and $\sigma_{{\delta f}^2}$ is important, because we have direct access to the first quantity, but not to the second.
This relation can be found for instance rewriting Eq. \ref{variance} in the case of Gaussian-distributed noise $\delta f$.
For simplicity, we define $\delta f = x$ and ${\delta f}^2 = y = x^2$, so that the Gaussian distribution of $\delta f$ can be written as
\begin{align}
\int_{-\infty}^{\infty} \frac{1}{\sqrt{2\pi}\sigma_{x}} e^{- \frac{{x}^2}{2 \sigma_{x}^2}} {\rm d}x=
\int_{0}^{\infty} \frac{2}{\sqrt{2\pi}\sigma_{x}} e^{- \frac{{x}^2}{2 \sigma_{x}^2}} {\rm d}x =
\int_{0}^{\infty} \frac{1}{\sqrt{2\pi}\sigma_{x}\sqrt{y}} e^{- \frac{y}{2\sigma_x^2}} {\rm d}y
\end{align}
where we performed the substitution $y=x^2$, so that ${\rm d}x = 1/{2\sqrt{y}} \, {\rm d}y$.
Notice that the domain of $y$ is ${\mathbb{R}^+}$, so we exploited the symmetry of the Gaussian distribution of $x$ to change the integration range from $\left(-\infty,+\infty\right)$ to $\left[0,+\infty\right)$.
The distribution of the variable $y = {\delta f}^2$ is thus $\frac{1}{\sqrt{2\pi}\sigma_x\sqrt{y}} e^{- \frac{y}{2\sigma_x^2}}$.
Its variance is by definition
\begin{align}
\sigma_y^2 =
\int_0^{\infty} \frac{1}{\sqrt{2\pi}\sigma_x\sqrt{y}} e^{- \frac{y}{2\sigma_x^2}} y^2 \, {\rm d}y - \left(
\int_0^{\infty} \frac{1}{\sqrt{2\pi}\sigma_x\sqrt{y}} e^{- \frac{y}{2\sigma_x^2}} y \, {\rm d}y
\right)^2 =
2 \sigma_x^4
\end{align}
This result tells us that $\sigma_{{\delta f}^2} = \sqrt{2} \sigma_{\delta f}^2$, so Eq. \ref{variance} becomes:
\begin{align}
\label{eq:gaussapproxsigma}
\sigma_{\Gamma}^2 = \alpha^2 \sigma_{\delta f}^2 + 2\beta^2\sigma_{\delta f}^4
\end{align}

Some important considerations can be done looking at our simple analytical model.
First, we expect different behaviors for different values of the finesse, since it is directly related to the cavity linewidth as well as the gain function and its derivatives.
In particular, we expect a higher sensitivity to the suppression effect for higher finesse.
The same holds for the width of the laser spectrum, since for a wider spectrum, more external teeth with a wider detuning are involved, contributing to the noise suppression.

At this point, a further question arises: is it possible to exploit $f_{\rm 0}$ and different finesse values to obtain a strongly suppressed intensity noise with a desired gain?
In the next Sections we will show that it is actually possible, though some experimental limitations occur.

\section{Simulations}\label{sec:simul}
The cavity-laser system has been simulated to comprehend the impact of $f_0$ (thus of $f_{\rm ceo}$) on the total gain, on its relative noise,  and on the peak power of the stored pulses.
Notice that, since we showed $\sigma_{P,{\rm rel}} =\sigma_{\Gamma,{\rm rel}}$, from now we will refer to it as a unique $\sigma_{\rm rel}$.
These simulations allow estimating the effective cost paid in terms of gain reduction to have a strong noise suppression.
All the calculations have been performed with Wolfram Mathematica and Matlab.
We took the experimental data as technical parameters for the simulations (see Sec. \ref{sec:experiment}), in particular, we set the repetition rate of the laser to $f_{\rm rep} = \SI{100}{\mega\hertz}$.
To simulate the spectrum shape $S\left(\nu\right)$, we used a supergaussian function of the $4^{\rm th}$ order of the form $S\left(\nu\right) \propto \exp\left[-\ln{2}\,\left(2\,\left(\nu-\nu_0\right)/{\rm \Delta\nu}\right)^8\right]$, where the FWHM $\Delta\nu$ is given by $\Delta\nu = -c/\lambda_0^2 \, \Delta\lambda$, and $\nu_0 = {c}/{\lambda_0}$, being $\Delta\lambda=\SI{2.7}{\nano\meter}$ the corresponding FWHM in the wavelength domain, and $\lambda_0=\SI{1035}{\nano\meter}$ the central wavelength.
As far as the cavity is concerned, we simulated a 4-mirrors crossed cavity in overcoupled configuration, switching between the two different values of the finesse of $4300$ and $1800$.
We set the offset of the PDH error signal to 0, assuming a perfect locking of mode $m_0$.
Nevertheless, values of $f_{\rm pdh}$ up to some \SI{}{\kilo\hertz} do not affect the results appreciably.
Then, we took $\sigma_{\delta f} = \SI{5}{\kilo\hertz}$, which is the value we experimentally measured.
Finally, we assumed $\delta f$ Gaussian distributed, so we estimated $\sigma_{\rm rel}$ from Eq. \ref{eq:gaussapproxsigma}.

\begin{figure}
\centering
\includegraphics[width=0.6\textwidth]{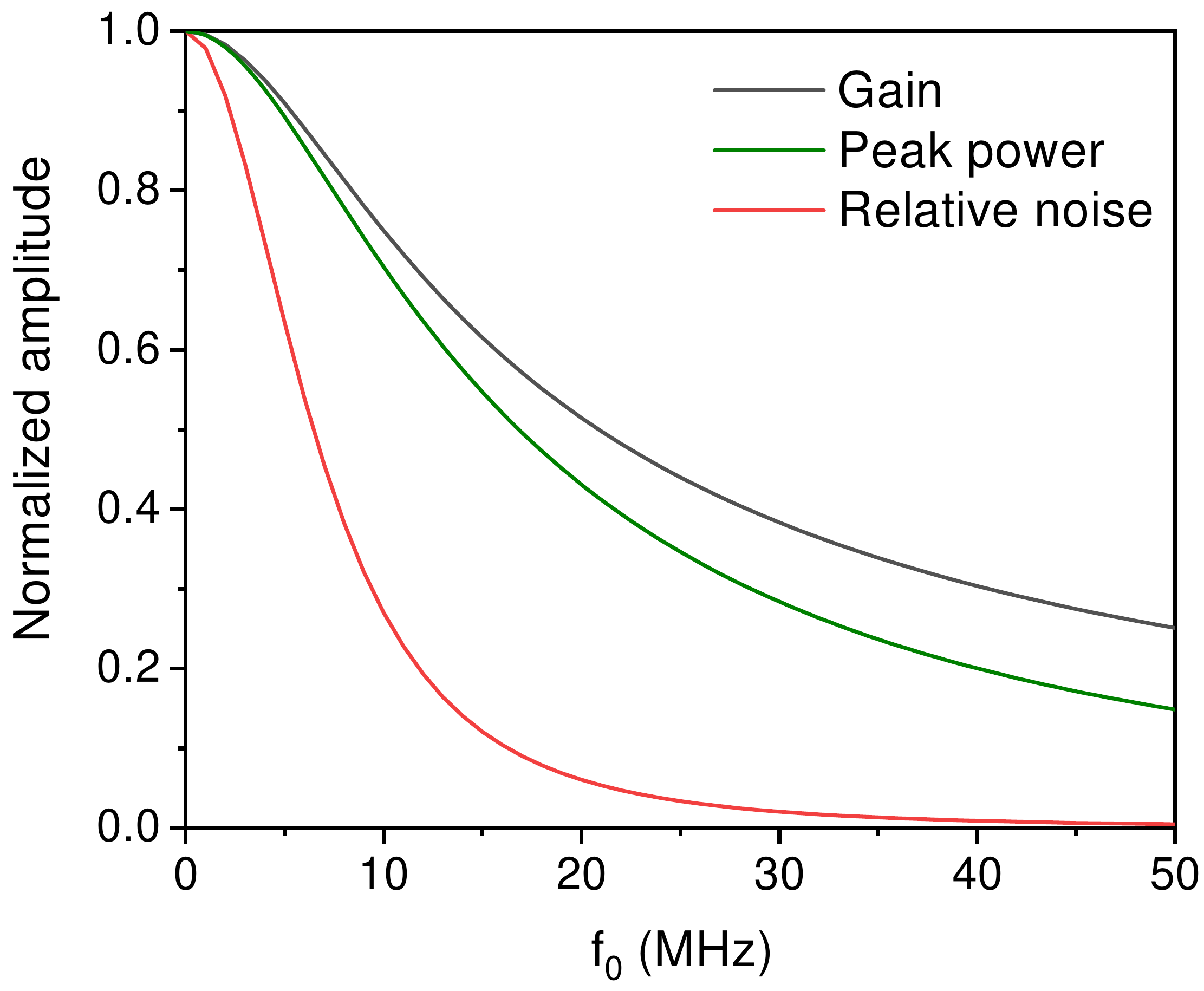}
\caption{From top to bottom: gain (black), stored pulses peak power (green), and relative power noise $\sigma_{\rm rel}$ (red) as functions of $f_0$.
Traces are normalized to their respective values in $f_{\rm off} = 0$.
Here the finesse is $4300$.
}
\label{simsample}
\end{figure}

\begin{figure}
\centering
\includegraphics[width=\textwidth]{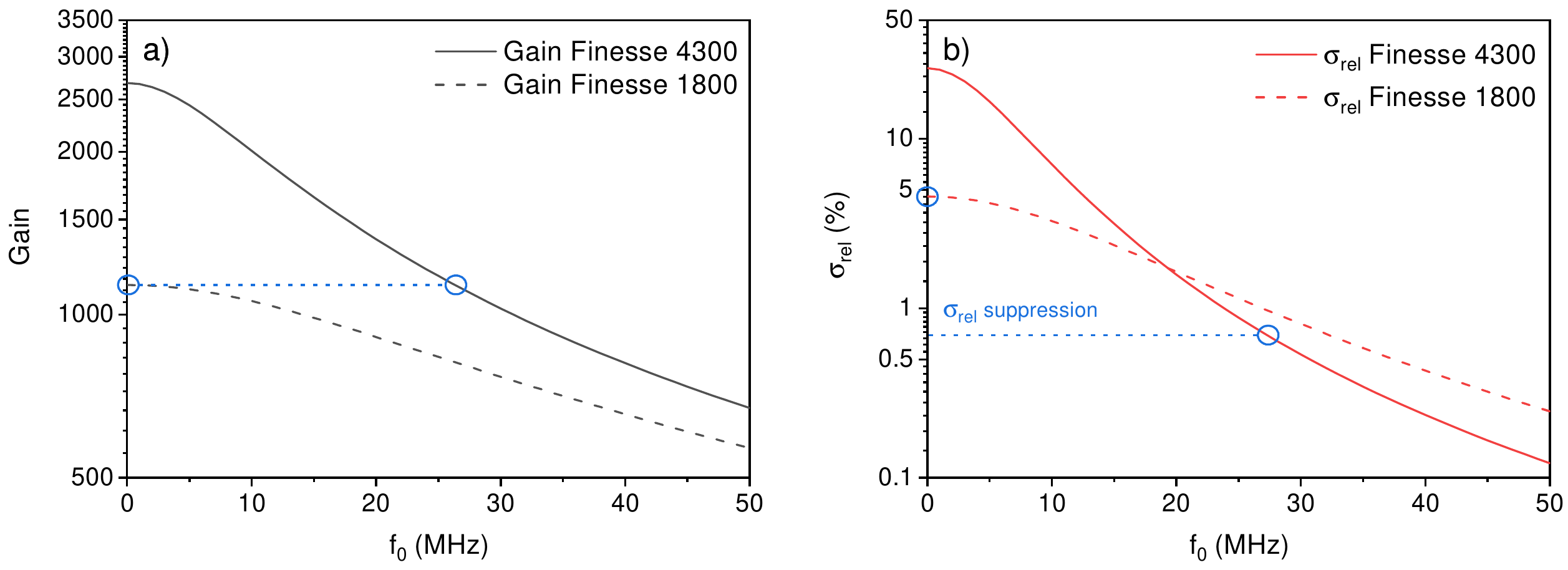}
\caption{Gain (a) and relative $\sigma_{\rm rel}$ (b) in logarithmic scale for $F=4300$ and $F=1800$.
For both configurations, the PDH offset is zero and $\sigma_{\delta f} = \SI{5}{\kilo\hertz}$.
The blue line in panel a) indicates the higher gain value possible for $F = 1800$ at $f_0=0$, namely gain $1130$.
The same gain is reached for finesse $4300$ at $f_0= \SI{30}{\mega\hertz}$.
In panel b), we show the suppression of $\sigma_{\rm rel}$ between the two configurations highlighted in panel a).}
\label{changingfinesse}
\end{figure}

Fig. \ref{simsample} shows the results of the simulations of normalized $\Gamma_{\rm tot}$, $\sigma_{\rm rel}$, and ${\rm P_{peak}}$ as functions of $f_0$, for a cavity of finesse $4300$.
We chose a normalized plot to highlight the different trends of the traces.
Notice that the peak power has been calculated from the laser spectrum, taking into account the cavity spectral filter effect and mode-dependent phase.
After multiplying all these terms, an Inverse Fourier Transform allows estimating the shape of the resulting pulses inside the cavity in temporal domain, thus $\rm P_{\rm peak}$.
%
%
As it can be easily noticed, all simulated quantities decrease when the laser-cavity offset $f_0$ rises.
Nevertheless, there is a faster drop in terms of relative noise with respect to the other traces.
For example, in the first $\SI{10}{\mega\hertz}$, the normalized amplitude of the noise reduces to $0.27$, while the peak power remains around $0.70$, and the total gain (average intracavity power) stays at about $0.75$.
Thus, in general, there is always an advantage in terms of signal-to-noise ratio for both $\rm P_{peak}$ and $\rm P_{cav}$, leaving the point $f_0 = 0$.
This general behavior has immediate repercussions in all those applications of non-linear optics such as the generation of high order harmonics (HHG) and ICS X-ray generation.
In HHG, a slight noise reduction in the fundamental harmonic could lead to a substantial enhancement for high-order harmonics noise.
Since the framework in which this work has been developed is the study and realization of optical cavities for ICS X-rays generation, we will concentrate only on the cavity gain and its noise, omitting further considerations on $\rm P_{peak}$.
Indeed, in ICS experiments, scattering efficiency is more sensitive to average power variations than to intracavity pulses temporal broadening \cite{zomerICS}.
As a second simulation, we compared the behaviors of two cavities with different values of finesse.
In particular, we calculated total gain and relative noise for finesse values of $4300$ and $1800$, and the results are reported in Fig. \ref{changingfinesse}.
In panel a) we report $\Gamma_{\rm tot}$, while in panel b) $\sigma_{\rm rel}$, for a cavity with finesse values of $4300$ and $1800$.
The gain decreases similarly for the two cases, although the maximum gain is a function of the finesse.
We start from a gain of $2680$ for $F=4300$, and of $1130$ for $F=1800$.
The noise at zero offset is different for the two cases, too, being \SI{26.2}{\percent} and \SI{4.6}{\percent}, respectively.
For both finesse values, the gain decreases slower than the noise.
From these simulations, the question arisen at the end of the theoretical section seems to have a positive answer: one can reduce the noise at a certain gain by setting a finesse higher than needed, and then increasing $f_0$ until the desired gain is reached, having a relative noise lower than the standard configuration of $f_0 = 0$.
For example, in this case the gain of $1130$ can be conveniently reached starting from $F = 4300$ and increasing $f_0$ until $\Gamma_{\rm tot} = 1130$ is reached (approximately $f_0 = \SI{27}{\mega\hertz}$, instead of choosing a finesse of $1800$ and couple it to the laser with $f_0=0$.
In this way, the same gain is achieved with a noise reduction of a factor $6.4$ ($\SI{0.72}{\percent}$ against $\SI{4.6}{\percent}$).

\section{Experiment}\label{sec:experiment}
\begin{figure}
\centering\includegraphics[width=0.7\textwidth]{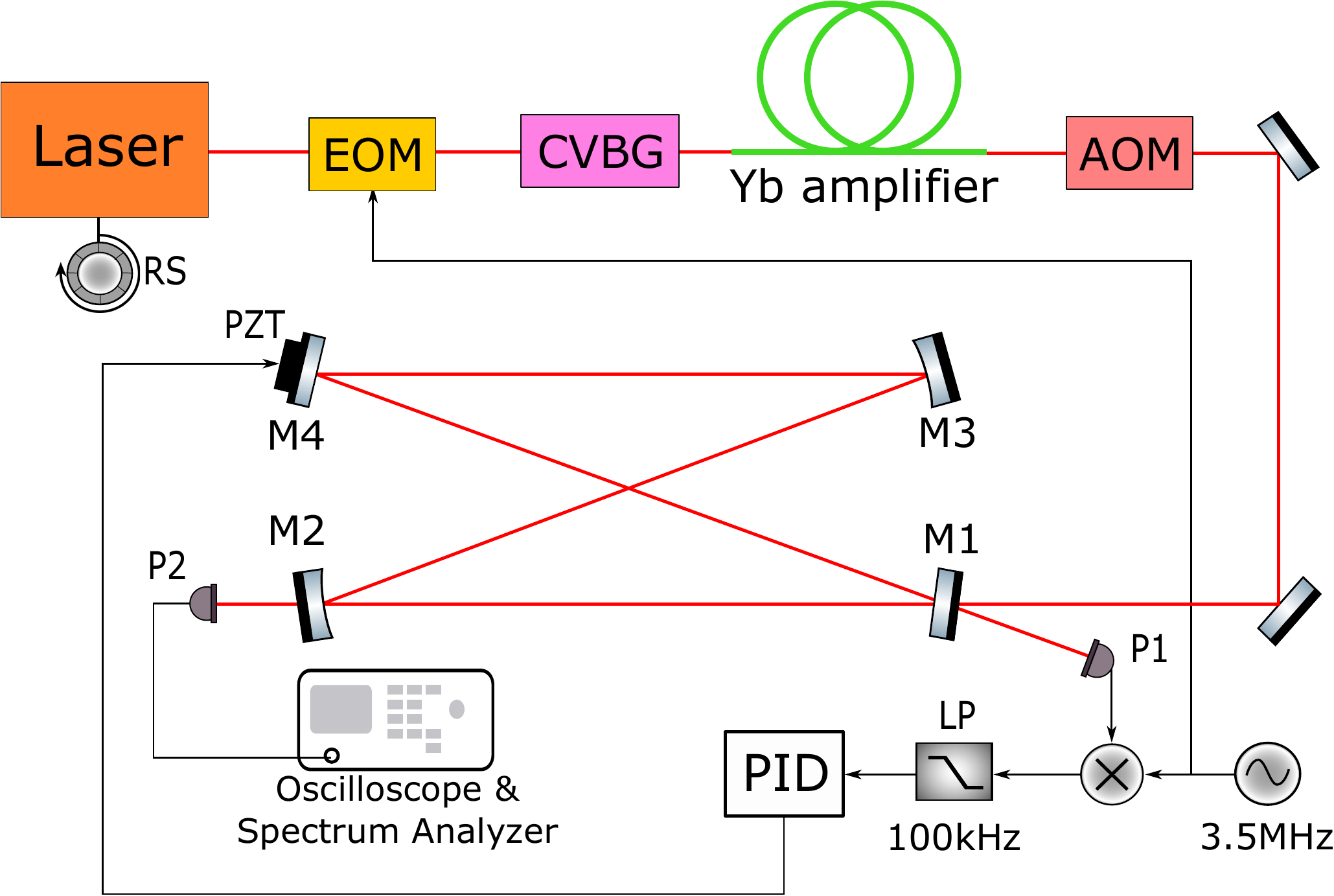}
\caption{Scheme of the setup used for our experiment.
The source laser is represented in orange.
RS: micrometric rotation stage; EOM: Electro-Optic modulator; CVBG: Chirped Volume Bragg Grating; AOM: Acousto-optical modulator; M1, M2, M3, M4: cavity mirrors; P1 and P2: photodetectors; LP: \SI{100}{\kilo\hertz} $5^{\rm th}$-order low-pass filter; PZT: piezoelectric actuator.}
\label{setup}
\end{figure}

In this Section we report on the experiment we performed to study and prove the $f_0$-dependent noise suppression.
The experimental setup is schematized in Fig. \ref{setup}.
The laser source is the Orange Yb:fiber mode-locked oscillator from Menlo Systems, with a repetition rate of \SI{100}{\mega\hertz}, and a bandwidth of \SI{20}{\nano\meter}, centered at \SI{1035}{\nano\meter}.
A BK7 window (with a thickness of \SI{5}{\milli\meter}) is inserted in the laser cavity, to control $f_{\rm ceo}$ by a manual micrometric rotation stage (labelled RS in the schematics).
The laser output power is on the order of \SI{200}{\milli\watt}.
The output pulses pass through a NewFocus Wideband 4004 IR Electro-Optic Modulator (EOM), that introduces a frequency modulation at \SI{3.5}{\mega\hertz}, needed for the Pound-Drever-Hall (PDH) cavity-laser stabilization.
Though not strictly required for this experiment (but already integrated in our experimental setup), an optical amplification stage follows.
The laser beam is stretched in time and selected in frequency domain by an Optigrate BG Pulse Chirped Volume Bragg Grating (CVBG) and power-enhanced by a \SI{4}{\meter} long Yb-fiber amplifier (based on Liekki Yb1200-12/125DC-PM), pumped by a multimode \SI{976}{\nano\meter} laser diode (Photontec M976).
The CVBG is required to avoid nonlinear effects inside the amplifier's fiber.
Here, the pulse length is stretched from \SI{200}{\femto\second} to \SI{380}{\pico\second}, while the spectrum is reduced to a FWHM of \SI{2.7}{\nano\meter}, with a shape well described by the $4^{\rm th}$-order supergaussian function used in the simulations.
The output power from the amplifier has been set to \SI{1.1}{\watt}.
Then the pulses go through a NEOS Acousto-optical Modulator (AOM), necessary for the measurement of the cavity finesse exploiting the technique described in \cite{accuratefinesse}.
At this point, the laser is coupled with a four-mirror crossed ring cavity.
The four mirrors by Layertec have a negligible dispersion in the spectral region of interest.
M1 and M4 are flat, while M2 and M3 are curved, with a radius of curvature of \SI{750}{\milli\meter}.
Since our cavity is overcoupled, we exploited two different input couplers M1 to switch between two different finesse values.
One input coupler has a power reflectivity of $R=\SI{99.66}{\percent}$ and gives a  finesse $1800$ (measured $1785\pm50$, $\Gamma_{\rm tot} = 1137\pm30$), while the other has $R=\SI{99.86}{\percent}$ and gives a finesse of $4300$ (measured $4270\pm110$, $\Gamma_{\rm tot} = 2720\pm70$).
All the other mirrors have a high reflectivity ($R>0.99999$).
The Free Spectral Range of the cavity is controlled by a piezoelectric actuator (PZT) attached to the mirror M4.
The active stabilization of the cavity against the laser is based on the PDH error signal generated from the beam reflected from M1.
It is detected by the photodetector P1 (Fermionics Opto-Technology FD500W), and low-pass filtered at \SI{100}{\kilo\hertz} to cancel the high-frequency components.
The PDH error signal is then sent to a PID controller, which elaborates the signal and applies it to the PZT.
The transmitted signal is measured by the photodiode P2 (Thorlabs PBD150A, bandwidth 5 MHz) behind M2.

\begin{figure}
    \centering
    \includegraphics[width=1\textwidth]{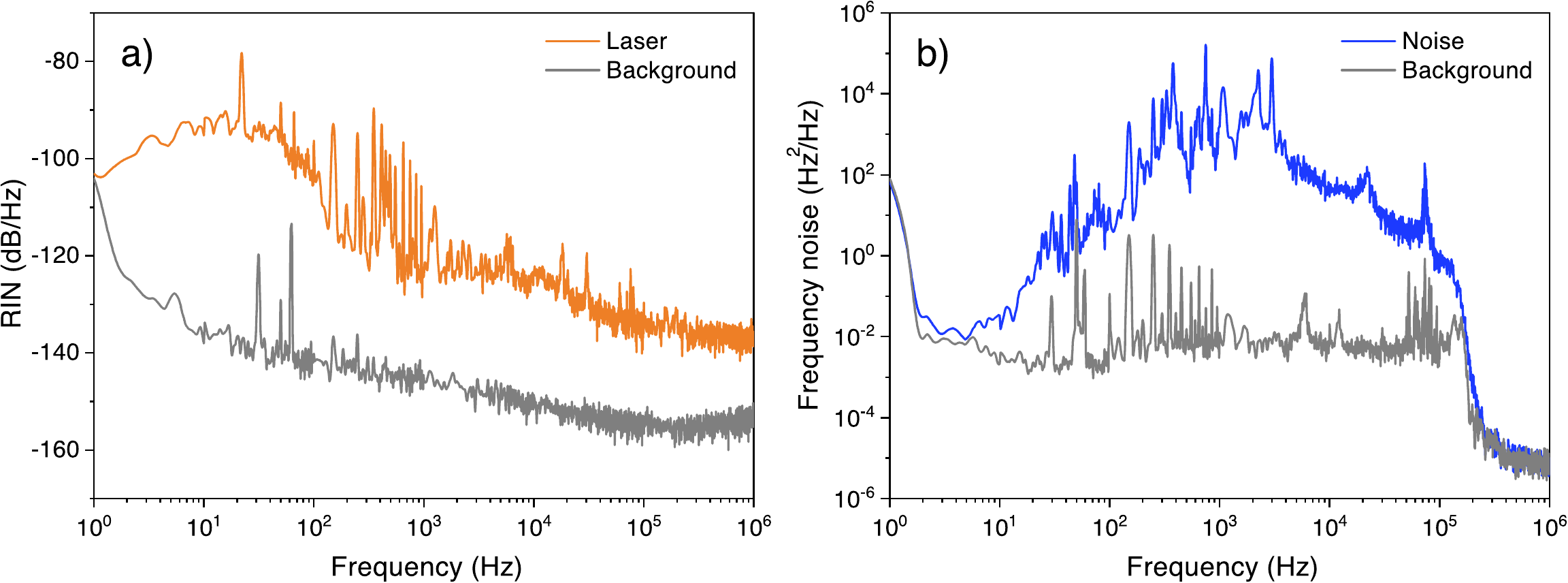}
    \caption{a) RIN of the laser before coupling to the cavity.
    b) Frequency noise of the laser-cavity coupled system.}
    \label{fig:RINlaserandfreq}
\end{figure}

We measured the Relative Intensity Noise (RIN) of the laser before coupling to the cavity, showing the results in the left panel of Fig. \ref{fig:RINlaserandfreq}.
The measurement has been performed using a large bandwidth photodetector and an Agilent E4445A Spectrum Analyzer, high-pass filtered at approximately \SI{150}{\hertz} to remove the DC component.
The high-pass (HP) filter response has been removed to estimate the integrated noises.
Then, the relative integrated noise from $\SI{1}{\hertz}$ to $\SI{1}{\mega\hertz}$ of the laser is $\sigma_{\rm rel} = \SI{0.03}{\percent}$.
For what concerns the RIN behavior, we observe a decreasing curve on the whole band (except for the region influenced by the filter).
The higher average level approaches \SI{-90}{\decibel\per\hertz}, except for a \SI{-80}{\decibel\per\hertz} noise peak, while the minimum approached at \SI{1}{\mega\hertz} is below \SI{-130}{\decibel\per\hertz}.
Several peaks are noticeable in the intensity noise spectrum, in particular between $\SI{100}{\hertz}$ and $\SI{2}{\kilo\hertz}$.
Those are due to the amplifier pump diode's electrical noise, which is directly transferred to the amplified signal, although it is cut at approximately $\SI{1}{\kilo\hertz}$ by Ytterbium spontaneous decay \cite{paschottayb}.

At this point, we coupled the laser to the cavity (in this case $F = 4300$), and we measured the frequency noise of the coupled system, shown in panel b) of Fig. \ref{fig:RINlaserandfreq}.
This measurement has been performed by acquiring the PDH error signal after LP, and converting it into a detuning signal by the so-called discriminator constant $k_{\rm d}$, defined as $k_{\rm d} = \frac{\delta \nu}{\delta V}$, where $\delta\nu$ is a detuning variation between the laser and the cavity frequencies, while $\delta V$ is the corresponding variation on the PDH signal.
We have $k_{\rm d} = \SI{1.56E5}{\hertz\per\volt}$.
To better estimate the noise without feedback contributions, such measurement has been performed with a weak lock.
The low-frequency peaks are mainly due to mechanical vibrations.
On the other hand, the noise around $\SI{10}{\kilo\hertz}$ comes from the piezoelectric actuator resonance.
Other contributions derives from the laser frequency noise spectrum but are not distinguishable from the cavity ones, since the discriminator measures only relative detunings.
The highest frequency noise level is reached by mechanical contributions, and it is approximately $10^3-10^4\SI{}{\hertz^2\per\hertz}$.
The sharp cut at \SI{100}{\kilo\hertz} is due to the LP filter of the PDH stabilization.
The integrated noise from panel b) gives the experimental $\sigma_{\delta f} = \SI{5063}{\hertz}$.

Exploiting the setup described above, we compared the Relative Intensity Noise of the signal transmitted from the mirror M2 with the two available finesse values and different $f_0$, bearing in mind that such signal is directly proportional to the intracavity power.
In addition, we acquired temporal traces of the transmitted beam to have a better visualization of the results.

\begin{figure}
\centering
\includegraphics[width=0.95\textwidth]{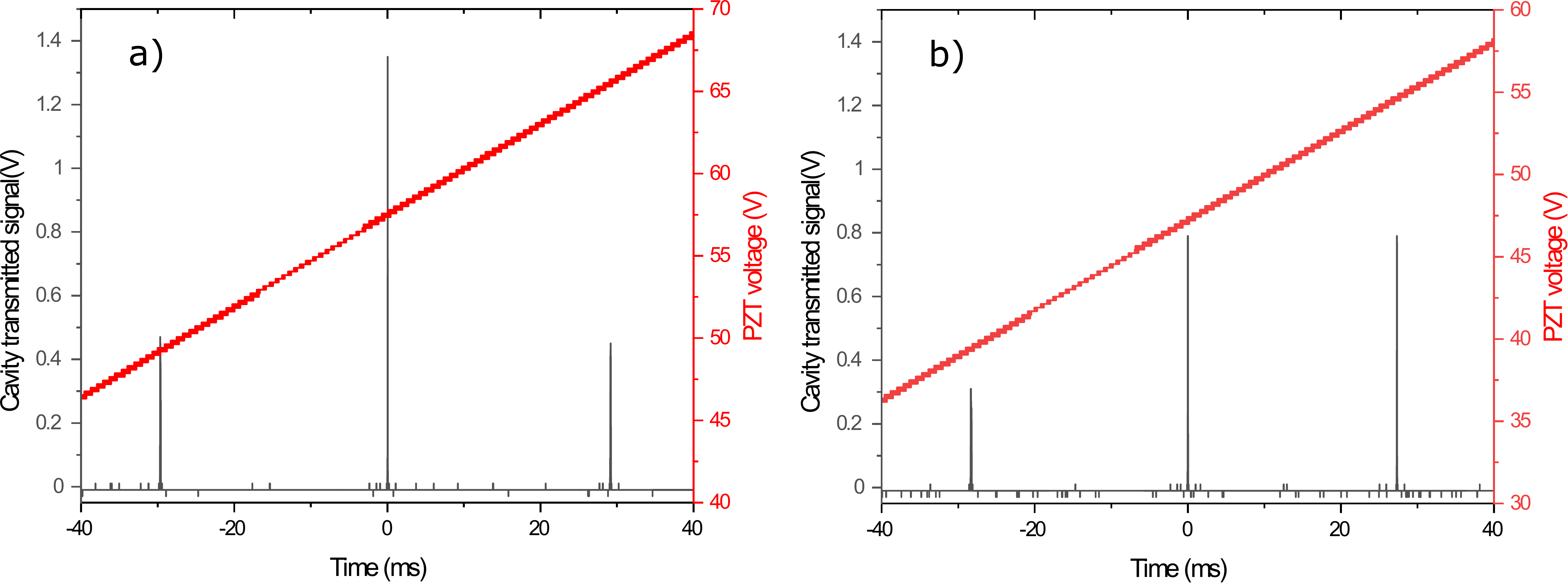}
\caption{Resonance peaks (black) and voltage applied to the piezoelectric actuator (red) during a scan of the cavity length.
The distance between two peaks is equal to a variation of the cavity length, corresponding to a $\rm FSR$.
In panel a), $f_0$ is null, thus the secondary peaks around the central one are symmetric and considerably lower than it.
In panel b) the opposite situation is shown: $f_0 = f_{\rm rep}/2 =\SI{50}{\mega\hertz}$, so the primary peak and its neighbor on the right have the same intensity and symmetry of secondary peaks is broken.}
\label{CEOpeaks}
\end{figure}

As mentioned at the beginning of this Section, the control of $f_0$ is allowed by the presence of a BK7 window inside the laser cavity.
The window can rotate at different angles with respect to the beam, thus modifying the intracavity dispersion, and inducing a $f_{\rm ceo}$ change.
As explained in Ref. \cite{gagliardi2014} and in Ref. \cite{Jones2002}, when $f_0$ is minimized, only one transmission peak is maximized in a cavity scan.
On the contrary, when $f_0 = f_{\rm rep}/2$ two consecutive peaks have the same intensity, visibly lower than the maximum of the previous situation.
We experimentally found these points (see Fig. \ref{CEOpeaks}) by adjusting the BK7 window angle, so calibrating the control of $f_0$.
Indeed, from simple goniometric considerations, $f_0$ can be written as a function of the window angle $\theta$:
\begin{equation}\label{ceowindow}
f_0\left(\theta\right) = 
\frac{f_{\rm rep}/2}{-\frac{1}{\sqrt{1-k^2\sin^2{\theta_0}}}+\frac{1}{\sqrt{1-k^2\sin^2{\theta_{\rm max}}}}}\left(\frac{1}{\sqrt{1-k^2\sin^2{\theta}}}-\frac{1}{\sqrt{1-k^2\sin^2{\theta_0}}}\right)
\end{equation}
where $\theta_0$ and $\theta_{\rm max}$ are the angular positions of $f_0=0$ and $f_0 = f_{\rm rep}/2$  respectively, taking the perpendicular position of the window as $\theta=0$, while $k = n_{\rm air}/n_{\rm BK7} = 1/1.507$ at \SI{1035}{\nano\meter}.
Estimating the uncertainties on the rotation stage and possible mount hysteresis, we claim an error of $\pm \SI{5}{\mega\hertz}$ on our $f_0$ measurements. 

\begin{figure}
\centering
\includegraphics[width=0.95\textwidth]{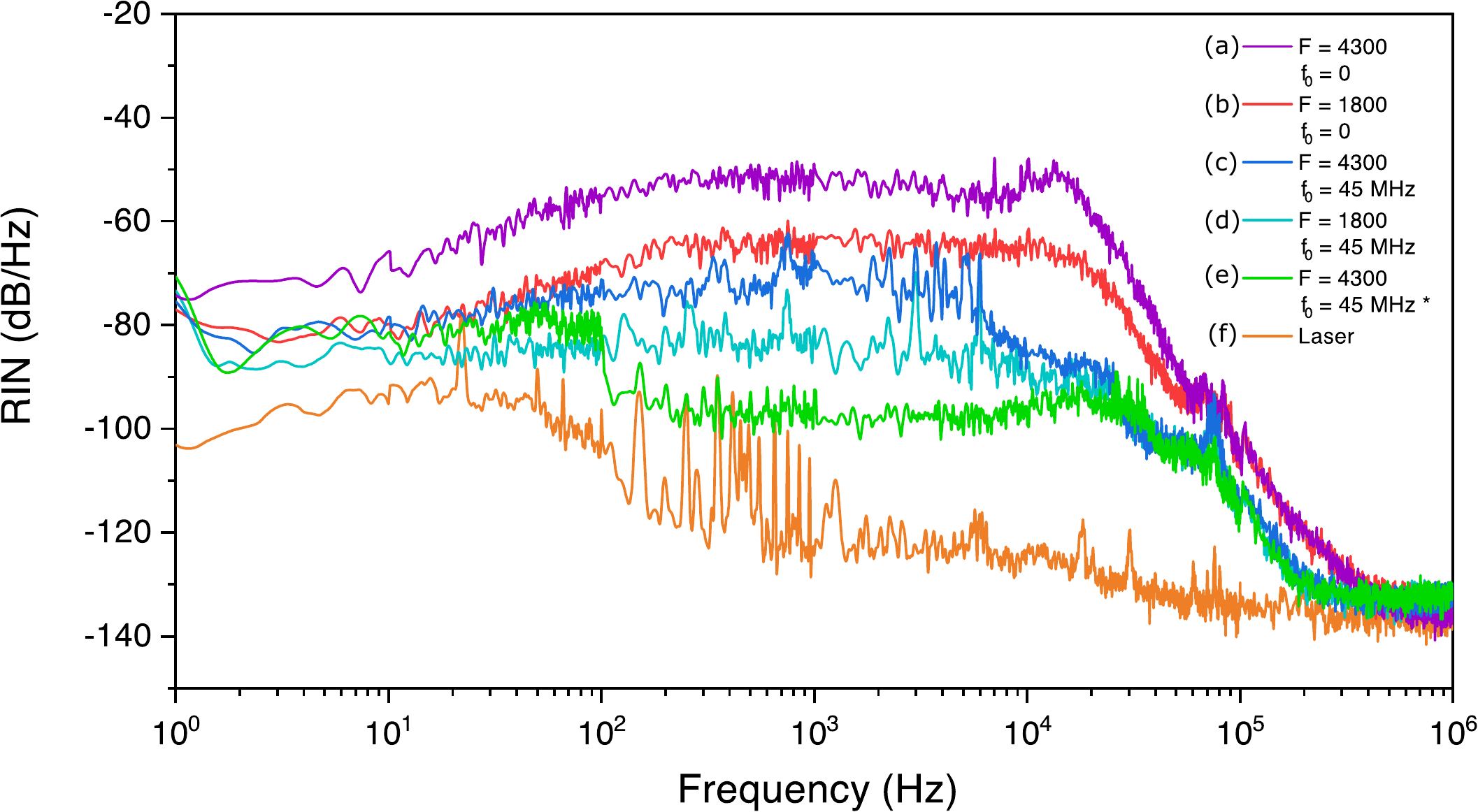}
\caption{RIN traces from photodetector P2.
From top to bottom (labeled in the legend both in colors and letters) traces acquired for: (a) $F = 4300$ and $f_0 = 0$, (b) $F = 1800$ and $f_0 = 0$, (c) $F = 4300$ and $f_0 = \SI{45}{\mega\hertz}$, (d) $F = 1800$ and $f_0 = \SI{45}{\mega\hertz}$, (e)$^*$ $F = 4300$ and $f_0 = \SI{45}{\mega\hertz}$ with PID parameters optimized for this case, and (f) laser RIN baseline.
Except for data (e) and (f), all the traces have been measured by leaving the PID cavity-lock parameters unaltered.}
\label{RIN}
\end{figure}

To investigate the noise transfer from the frequency detuning to the stored power, we acquired different RIN traces, exposed in Fig. \ref{RIN}.
Here, a comparison between the transmitted power RIN for finesse $1800$ and for finesse $4300$ in both cases with $f_0=0$ and $f_0=\SI{45}{\mega\hertz}$ is shown (curves from (a) to (d)).
As a reference baseline for these measurements, we report the laser RIN in orange (f).
The choice of $f_0 = \SI{45}{\mega\hertz}$ is not accidental, but we experimentally set it to the value such that the configuration with $F=4300$ and $f_0=\SI{45}{\mega\hertz}$ has the same gain as the configuration with $F=1800$ and $f_0=0$.
Indeed, we measured the same cavity passive gain comparing the transmission power levels for finesse $1800$, $f_0=0$ and finesse $4300$, $f_0 = \SI{45}{\mega\hertz}$, so having the occasion to directly compare the contribution of $f_0$ on noise suppression.
It is worth noting that we fixed the PID lock parameters (chosen by optimizing the cavity lock for finesse $1800$, $f_0=0$) in order to compare different finesse values and offsets without changing the loop gain or bandwidth.
As a last step, we acquired also a RIN of the transmitted signal with finesse $4300$ and $f_0 = \SI{45}{\mega\hertz}$ optimizing the feedback parameters (trace (e)).
This latter measurement shows a further enhancement in terms of power fluctuations.
The higher trace, namely the one labeled (a), is the one of finesse $4300$ and $f_0 = 0$.
Before \SI{10}{\kilo\hertz} it is constant at about \SI{-50}{\decibel\per\hertz}, while a broad peak at \SI{11}{\kilo\hertz} appears at the piezoelectric actuator resonance.
After the resonance, the trace falls and reaches the background floor around \SI{400}{\kilo\hertz}.
The same drop is noticeable in trace (b), which is the RIN of finesse $1800$ and $f_0=0$.
On the other hand, the average is \SI{20}{\decibel} lower than the previous, and no piezoelectric actuator resonances are visible.
A substantial change is noticeable looking at trace (c), acquired for finesse $4300$ but high $f_0$.
Though, the gain is the same for this curve and (b), while the noise is on average \SI{10}{\decibel} lower between \SI{100}{\hertz} and \SI{10}{\kilo\hertz}, and \SI{20}{\decibel} lower between \SI{10}{\kilo\hertz} and \SI{50}{\kilo\hertz}.
Furthermore, the noise fall reaches the floor at \SI{200}{\kilo\hertz} instead of \SI{400}{\kilo\hertz}.
A similar behavior can be encountered for trace (d), namely the one obtained from finesse $1800$ and $f_0 = \SI{45}{\mega\hertz}$ (with a lower average of about \SI{10}{\decibel} before \SI{10}{\kilo\hertz}).
A difference in the RIN fall around \SI{10}{\kilo\hertz} of the two curves (a) and (b) and the three curves (c), (d), and (e) can be observed.
We attribute it to the general noise reduction in presence or absence of $f_0$, thus a rescaling of the RIN level also for frequencies above the PID bandwidth.
The best noise suppression case is then represented in trace (e), where we combined the effect of $f_0$ with an optimization of the feedback parameters.
The feedback noise reduction is evident between \SI{100}{\hertz} and \SI{10}{\kilo\hertz} (which is the feedback bandwidth, i.e., the actuator mechanical resonance frequency), where the noise remains on average below \SI{-90}{\decibel\per\hertz}.

From the RINs, we calculated the relative noise integrating from \SI{1}{\hertz} to \SI{1}{\mega\hertz} obtaining the following values:
$\sigma_{\rm a} = \SI{32.5}{\percent}$, $\sigma_{\rm b} = \SI{8.0}{\percent}$, $\sigma_{\rm c} = \SI{2.5}{\percent}$, $\sigma_{\rm d} = \SI{1.2}{\percent}$, and $\sigma_{\rm e} = \SI{0.4}{\percent}$.
Sigmas are labeled with the same letter of the correspondent RIN. Notice that there is good correspondence between the simulated noise decreasing trend and the experimental values.

The core of this work resides in the difference between $\sigma_{\rm b}$ and $\sigma_{\rm c}$.
Indeed, these measurements have been obtained at the same gain (thus at the same cavity stored power), but the power fluctuations are lowered by $f_0$ of a factor $3.2$.
Of course, the compensation becomes more evident after the PID optimization, which leads to an integrated relative noise reduced of a factor $20$ with respect to the one from trace (b).

\begin{figure}
\centering
\includegraphics[width=0.6\textwidth]{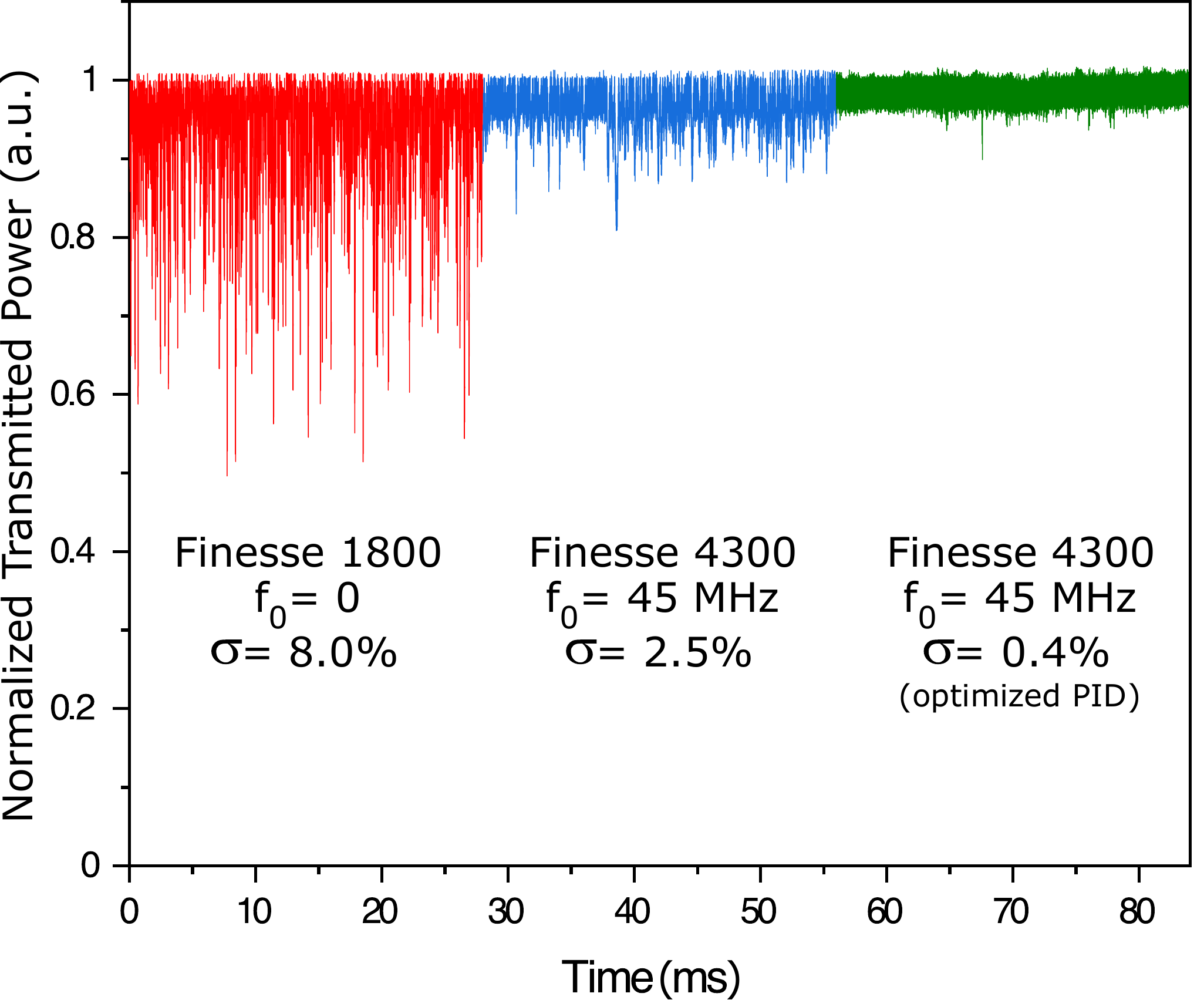}
\caption{Time domain traces of finesse $1800$ - $f_0 = 0$ (red, left), finesse $4300$ - $f_0 = \SI{45}{\mega\hertz}$ before (blue, central) and after (green, right) the PID optimization.
Cavity gain (thus stored power) is essentially the same in all represented cases.
The traces have been merged in a unique time axis for simplicity of representation.}
\label{trasmittedpower}
\end{figure}

For a better visualization of the results, we report the three cases of interest (b), (c), and (e) transmission power traces in Fig. \ref{trasmittedpower}.
All the represented measures were acquired by the same detector used for the RINs and sent to the oscilloscope.
The color used in the picture recalls the correspondent cases in Fig. \ref{RIN}. Thus, from left to the right, we encounter cases (b), (c), and (e).
As expected, the relative noise is subjected to a drop as $f_0$ increases, passing from $\sigma_{\rm b} = \SI{8.0}{\percent}$ for $F=1800$ and $f_0=0$, to $\sigma_{\rm e} = \SI{0.4}{\percent}$ for $F=4300$ and $f_0 = \SI{45}{\mega\hertz}$ and optimized feedback, while maintaining the same gain.
Notice that even though the order of magnitude of the relative noises is the same of simulation (as well as their drop at increasing $f_0$), the measured values are lower than the predicted ones in most of the cases.
This fact can be ascribed to a partial suppression given by the PDH feedback, as can be seen from the RIN traces.

A last comment should be addressed to the comparison between the simulations and the experimental data.
Indeed, the value of $f_0$ at which the two configurations have the same gain is quite different between the simulations and the experiment (\SI{30}{\mega\hertz} versus \SI{45}{\mega\hertz}, respectively).
This might be due to an approximated modeling of the spectrum and to uncertainties on the $f_0$ measurement.
Actually, this is not the point, since the aim of this work is not to find the exact values of the gain and the noise as functions of $f_0$, rather to prove that gain drops slower than relative noise, and that this behavior can be effectively exploited in real setups.

\section{Conclusions}\label{sec:concl}
In conclusion, we demonstrated both theoretically and experimentally the suppression effect of $f_0$ (thus of $f_{\rm ceo}$) over the frequency noise contribution to power fluctuations of a laser-cavity locked system.
We showed that increasing $f_0$ leads to a substantial noise reduction, visible both in terms of integrated relative noise and its spectrum.
We also demonstrated that the cavity gain decreases slower than the noise.
This finding opens the possibility of exploiting higher finesse to obtain a desired cavity gain (stored power), while maintaining substantially lower power instabilities.
In particular, we experimentally showed that a cavity gain of approximately $\Gamma_{\rm tot} = 1130$ can be obtained either with a finesse of $1800$ and an offset $f_0=0$, or with a finesse of $4300$ and an offset $f_0 = \SI{45}{\mega\hertz}$.
However, the relative noise in these two cases is very different, passing from \SI{8.0}{\percent} to \SI{2.5}{\percent}, respectively.
In the last configuration, an optimization of the PID parameters allowed us to further decrease the relative noise to \SI{0.4}{\percent}.



\end{document}